\documentclass{elsart}
\usepackage{graphicx}
\usepackage{amsmath}
\usepackage{amssymb}
\usepackage{epsfig}
\usepackage{dcolumn}
\usepackage{bm}
\begin{document}
\begin{frontmatter}

\title{Generalized thermostatistics \\ based on multifractal phase space}
\author{A.I. Olemskoi}
\ead{alex@ufn.ru}
\address{Department of Physical Electronics,
Sumy State University,\\ 2, Rimskii-Korsakov St., 40007 Sumy, Ukraine}
\date{}

\begin{abstract}
We consider the self-similar phase space with reduced fractal dimension $d$
being distributed within domain $0<d<1$ with spectrum $f(d)$. Related
thermostatistics is shown to be governed by the Tsallis' formalism of the
non-extensive statistics, where role of the non-additivity parameter plays
inverted value ${\bar\tau}(q)\equiv 1/\tau(q)>1$ of the multifractal function
$\tau(q)= qd(q)-f(d(q))$, being the specific heat, $q\in(1,\infty)$ is
multifractal parameter. In this way, the equipartition law is shown to take
place. Optimization of the multifractal spectrum $f(d)$ derives the relation
between the statistical weight and the system complexity.
\end{abstract}

\end{frontmatter}

\section{Introduction}

Generalization of the thermostatistics is known to be based on the deformation
procedure of both logarithm and exponential \cite{Tsallis,Naudts,1}.
The simplest way to introduce these functions into the thermostatistics scheme
is to consider the equation of motion for dimensionless volume
$\gamma=\Gamma/(2\pi\hbar)^{6N}$ of the supported phase space ($\hbar$, $N$
being Dirac-Planck constant and particle number). In the course of evolution of
the ensemble with statistical weight $w=w(\gamma)$ and entropy $S$, the
variation rate of the phase space volume is governed by the following equation
\cite{1}
\begin{equation}
\frac{{\rm d}\gamma}{{\rm d}t}=w(\gamma)\frac{{\rm d}S}{{\rm d}t}.\label{1}
\end{equation}
From this, one follows the relation ${\rm d}S={\rm d}\gamma/w(\gamma)$ whose
integration gives the entropy related to the whole statistical weight $W$:
\begin{equation}
S(W)=\int\limits_{\gamma(1)}^{\gamma(W)}\frac{{\rm
d}\gamma}{w(\gamma)}.\label{2}
\end{equation}
Here, we take into account that entropy of a single state $S(w=1)$ vanishes.

In the case of the smooth phase space, one has trivial relation
$w(\gamma)=\gamma$ whose insertion into the relation (\ref{2}) derives to the
Boltzmann entropy $S=\ln W$. However, complex systems has fractal phase space
with the dimension $D<6N$, so that relation between the statistical weight and
the corresponding volume should be replaced by the power-law dependence
\begin{equation}
w(\gamma)=\gamma^d/d \label{3}
\end{equation}
where reduced fractal dimension $d\equiv{D}/{6N}\leq 1$ is introduced.
Insertion of Eq. (\ref{3}) into integral (\ref{2}) gives the expression
\begin{equation}
S(W)=\ln_{\bar d}\left(dW\right),\quad\ln_{\bar d}(x)\equiv\frac{x^{{\bar
d}-1}-1}{{\bar d}-1}\label{4}
\end{equation}
which is reduced to the Tsallis' logarithm $\ln_q(x)$ where non-additivity
parameter $q$ is replaced by inverted value ${\bar d}\equiv 1/d\geq 1$ of the
reduced fractal dimension $d$ of the phase space. Naturally, this expression is
reduced to the Boltzmann entropy in the limit $d\to 1$.

Adduced formalism is based on the proposition that the phase space is a
monofractal determined by single dimension $d$. The aim of this letter is to
generalize Tsallis' thermostatistics onto the multifractal phase space with a
spectrum $f(d)$. Such a generalization for arbitrary distribution $f(d)$ is
carried out in Section 2. Discussion in Section 3 shows that overall
representation of the thermostatistics based on the multifractal phase space
demands of the consistent consideration of input, escort and `physical'
distributions. An optimization of the spectrum $f(d)$ is considered in Section
4. Section 5 concludes our treatment.

\section{Multidractal phase space}

According to the usual recipe \cite{Feder}, the passage to multidractal phase
space is provided with replacement of the power function (\ref{3}) by the
expression $\gamma^{qd-f(d)}$ where $q$ is multifractal index, multiplier
$\gamma^{-f(d)}$ takes into account specific number of monofractals with
dimension $d$ within the multifractal. As a result, the statistical weight,
playing the role of the multifractal measure, takes the form
\begin{equation}
w_q(\gamma)=\int\limits^1_0\gamma^{qd-f(d)}\rho(d){\rm d}d \label{8}
\end{equation}
where $\rho(d)$ is the density distribution over dimensions $d$. Using the
method of steepest descent, we arrive at the modified power law
\begin{equation}
w_q(\gamma)\simeq\gamma^{\tau(q)}\label{9}
\end{equation}
which generalizes the simplest relation (\ref{3}) due to replacement of the
bare fractal dimension $d$ by the multifractal function $\tau(q)=
qd(q)-f(d(q))$. The conditions of application of the steepest descent method
\begin{equation}
\left.\frac{{\rm d}f}{{\rm d}d}\right|_{d=d(q)}=q,\quad\left.\frac{{\rm d}^2
f}{{\rm d}d^2}\right|_{d=d(q)}<0 \label{10}
\end{equation}
allow us to find special fractal dimension $d(q)$ related to given
parameter $q$.

Above consideration shows that the passage from monofractal phase space to
multifractal one is obtained by replacement of the single dimension $d$ by the
function $\tau(q)$ that monotonically increases, taking value $\tau=-1$ at
$q=0$ and $\tau=0$ at $q=1$. Limit behaviour of the function $\tau(q)$ is
characterized by the asymptotics
\begin{equation}
\tau\propto(q-1)\ \ \text{at}\ \ 0<q-1\ll 1,\qquad\tau\simeq 1\ \ \text{at}\ \
q\to\infty. \label{11}
\end{equation}
Physical domain of the $q$ parameter variation is bounded by condition $q>1$
which ensures positive values of the function $\tau(q)<1$.

As a result, we can use well-known Tsallis' formalism of the non-extensive
statistics where role of the non-additivity parameter plays inverted value
${\bar\tau}(q)\equiv 1/\tau(q)>1$ of the multifractal function $\tau(q)=
qd(q)-f(d(q))$. Thus, the entropy in dependence of the probability distribution
$ P_i$ has the form \cite{Tsallis}
\begin{equation}
S_q=-\sum\limits_{i=1}^{W_q} P_i\ln_{{\bar\tau}(q)}( P_i),\quad
\ln_{\bar\tau}(x)\equiv\frac{x^{{\bar\tau}-1}-1}{{\bar\tau}-1}. \label{12}
\end{equation}
With accounting the conditions
\begin{equation}
\sum\limits_{i=1}^{W_q} P_i=1,\quad E_q=\sum\limits_{i=1}^{W_q}\varepsilon_i
P_i^{{\bar\tau}(q)},\label{13}
\end{equation}
this expression arrives at the generalized distribution over energy levels
$\varepsilon_i$:
\begin{equation}
 P_i=Z_q^{-1}\exp_{{\bar\tau}(q)}\left(-\beta\varepsilon_i\right)
\label{14}
\end{equation}
where $Z_q$ is the partition function and the deformed exponential is
determined in the usual manner:
\begin{eqnarray}
\exp_{\bar\tau}(x)\equiv\left\{
\begin{array}{ll}
\left[1+({\bar\tau}-1)x\right]^{\frac{1}{{\bar\tau}-1}}\ {\rm at}\
1+({\bar\tau}-1)x>0,\\ 0\ \ \qquad\quad\qquad\qquad\quad\qquad\quad{\rm
otherwise}.
\end{array} \right.
\label{D}
\end{eqnarray}
The normalization condition (\ref{13}) gives the relation
\begin{equation}
\sum_{i=1}^{W_q}
P_i^{{\bar\tau}(q)}=Z_q^{-\left[{\bar\tau}(q)-1\right]}-\left[{\bar\tau}(q)-1\right]\beta
E_q. \label{15}
\end{equation}

In spite of the formal similarity of above expressions with the same of the
Tsallis' statistics, a principle difference takes place: in the last case, the
limit of extensive systems is reached, when non-additivity parameter takes
value $q=1$, whereas the multifractal phase space is reduced to monofractal one
with dimension ${\bar\tau}(q)\to 1$ in the opposite limit $q\to\infty$. In this
way, the step-function spectrum $\tau(q)$, being $\tau=1$ at $q>1$ and $\tau=0$
otherwise, relates to the smooth phase space.

Thermodynamic functions of the model under consideration can be found
analogously to the Tsallis' non-extensive scheme \cite{Tsallis}. However,
related expressions are very cumbersome even in the simplest case of the ideal
gas \cite{3,4,5} and amount to the usual form only within the slightly
non-extensive limit \cite{6}. At the same time, developed scheme allows one to
use thermodynamic formalism of multifractal objects \cite{BS}. Within the
latter, the role of a state parameter plays the multifractal index $q$, whose
variation may arrive at phase transitions if the dependence $\tau(q)$ has some
singularities. It is worthwhile to stress that developed scheme arrives
directly at related singularities of thermodynamic functions type of the
internal energy (\ref{18}) (see below), the entropy (cf. Eq.(\ref{4}))
\begin{equation}
S_q={\bar\tau}(q)\ln_{{\bar\tau}(q)}\left(W_q\right),\quad{\bar\tau}(q)\equiv
1/\tau(q)\label{4a}
\end{equation}
and the free energy
\begin{equation}
F_q=E_q-TS_q.\label{4b}
\end{equation}

\section{Discussion}

In accordance with the non-extensive thermostatistics \cite{Tsallis,3} the
distribution (\ref{14}) plays the role of the escort probability related to the
input one
\begin{equation}
p_i\equiv P^{{\bar\tau}(q)}_i=Z_q^{-{\bar\tau}(q)}
\left[\exp_{{\bar\tau}(q)}\left(-\beta\varepsilon_i\right)\right]^{{\bar\tau}(q)},
\label{26}
\end{equation}
being normalized by the condition
\begin{equation}
\sum\limits_{i=1}^{W_q}p_i^{\tau(q)}=1. \label{27}
\end{equation}
Easily to convince, the expression (\ref{26}) can be reduced to pseudo-Gibbs
form
\begin{equation}
p_i=\exp_\tau\left[\beta_q\left(F_q-\varepsilon_i\right)\right]\label{28}
\end{equation}
if one introduces effective value of the inverted temperature
\begin{equation}
\beta_q\equiv{\bar\tau}(q)Z_q^{-\left[{\bar\tau}(q)-1\right]}\beta\label{29}
\end{equation}
and determines the free energy $F_q$ in the usual manner:
\begin{equation}
F_q\equiv-T\ln_{{\bar\tau}(q)}(Z_q). \label{30}
\end{equation}
It is characteristically, the escort distribution (\ref{14}), determined with
the physical temperature $T\equiv\beta^{-1}$, is normalized by the standard
condition (\ref{13}), whereas the input probability (\ref{26}), being deformed
exponential (\ref{28}), is determined with non-physical temperature (\ref{29})
to derive to the deformed free energy (\ref{30}). Thus, both input and escort
distributions appear as complementary ones. In this way, the escort
distribution (\ref{14}) is deformed with the inverted index ${\bar\tau}(q)$,
whereas the input probability (\ref{28}) -- with the index $\tau(q)$ itself.
This exhibits the known `$\tau(q)\leftrightarrow{\bar\tau}(q)$'-duality of
Tsallis' thermostatistics \cite{Naudts}.

According to \cite{3} the physical distribution is neither escort nor
input probabilities, but the following one
\begin{equation}
\mathcal{P}(\varepsilon_i)\equiv\frac{P^{{\bar\tau}(q)}(\varepsilon_i)}
{\sum_{i=1}^{W_q}P^{{\bar\tau}(q)}(\varepsilon_i)}. \label{a}
\end{equation}
It corresponds to the condition
\begin{equation}
\sum\limits_{i=1}^{W_q}(\varepsilon_i-E_q)P^{{\bar\tau}(q)}(\varepsilon_i)=0
\label{b}
\end{equation}
instead of the second equation (\ref{13}). Easily to show, the
probabilities (\ref{26}), (\ref{a}) are connected by the relation
\begin{equation}
\mathcal{P}(\varepsilon_i)=Z_q^{\left[{\bar\tau}(q)-1\right]}p(\varepsilon_i-E_q)
\label{cc}
\end{equation}
to derive the distribution
\begin{equation}
\mathcal{P}(\varepsilon_i)=Z_q^{-1}\exp_{\tau(q)}\left[-{\bar\tau}(q)\beta
\left(\varepsilon_i-E_q\right)\right]. \label{c}
\end{equation}

In the case of continuous energy spectrum characterized with the density
distribution $\rho(\varepsilon)$, the internal energy related to the condition
(\ref{b}) takes the form
\begin{equation}
E_q=\int\limits_{-\infty}^{\infty}\varepsilon
\mathcal{P}(\varepsilon)\rho(\varepsilon){\rm d}\varepsilon. \label{16}
\end{equation}
Extreme value of $E_q$ is reached at the condition
\begin{equation}
\frac{\rho^{'}(\varepsilon)} {\rho(\varepsilon)}\simeq-\frac{
\mathcal{P}^{'}(\varepsilon)}{\mathcal{P}(\varepsilon)}
\label{17}
\end{equation}
where prime denotes differentiation over $\varepsilon$. Usually, the density
function is reduced to the power law $\rho(\varepsilon)\sim\varepsilon^{cN}$,
$c\sim 1$, so that $\rho^{'}(\varepsilon)/\rho(\varepsilon)\simeq
cN/\varepsilon$. Then, with using the distribution (\ref{c}), the condition
(\ref{17}) taken at $\varepsilon=E_q$ arrives at the equipartition law
\begin{equation}
E_q=c\tau(q)NT\label{18}
\end{equation}
according to which the value $c\tau(q)$ is the specific heat.

Let us stress that above scheme is related to generalized definition of the
deformed logarithm \cite{Naudts}
\begin{equation}
\ln_\phi(x)\equiv\int\limits_1^x\frac{{\rm d}y}{\phi(y)} \label{31a}
\end{equation}
where the function $\phi(y)$ is reduced to the statistical weight distribution
$w_q=p_i^{\tau(q)}$ type of Eq.(\ref{9}). However, this function is obeyed the
condition $p_i{'}=-\beta_q w_q(p_i)$ which includes deformed temperature
$\beta_q^{-1}$, being inconvenient at physical considerations.

\section{Optimization of multifractal spectrum}

Up to now, we suppose that the multifractal spectrum $f(d)$ is arbitrary. If it
is optimized at normalization condition
\begin{equation}
\int\limits_0^1 f(d){\rm d}d=1, \label{19}
\end{equation}
one has to minimize the expression
\begin{equation}
\tilde{S}_q\{f(d)\}=\int\limits^{\gamma(W_q)}_{\gamma(1)}\left[\int\limits^1_0\gamma^{qd-f(d)}\rho(d){\rm
d}d\right]^{-1}{\rm d}\gamma-\frac{\Sigma ^2}{2}\left[\int\limits_0^1 f(d){\rm
d}d-1\right], \label{20}
\end{equation}
written with accounting Eqs.(\ref{2}), (\ref{8}), where $\Sigma $ is Lagrange
multiplier. As a result, we arrive at the equality
\begin{equation}
\int\limits^{\gamma(W_q)}_{\gamma(1)}\gamma^{\left\{\left[qd-f(d)\right]-2\tau(q)\right\}}\ln\gamma~{\rm
d}\gamma=\frac{\Sigma ^2}{2}\label{21}
\end{equation}
whose integration gives, with accounting Eq.(\ref{9}), the transcendental
equation
\begin{eqnarray}
\frac{1}{2}\left[\Sigma\tau(q){\mathcal T}_q(d)\right]^2-W_q^{{\mathcal T}
_q(d)}\left[{\mathcal T}_q(d)\ln(W_q)-1\right]-1=0,\label{22}\\ {\mathcal
T}_q(d)\equiv \bar{\tau}(q)\left\{1-2\tau(q)+\left[qd-f(d)\right]\right\}.
\label{23}
\end{eqnarray}
This equation is written in the form, when can be used either given spectrum
function $f(d)$ or the index dependence $\tau(q)$. In the latter case, we find
initially the dependence $q(d)$ from the equation
\begin{equation}
\left.\frac{{\rm d}\tau}{{\rm d}q}\right|_{q=q(d)}=d,
\label{10a}
\end{equation}
being conjugated to Eq.(\ref{10}). Then, substituting this dependence into
Eq.(\ref{23}), we arrive at the trivial expression
\begin{equation}
{\mathcal T}(q)\equiv{\mathcal T}_q(d(q))=\bar{\tau}(q)-1\label{23a}
\end{equation}
whose using in Eq.(\ref{22}) allows us to determine the dependence of the
statistical weight $W_q$ on the complexity $\Sigma$ at given function
$\tau(q)$.

With passage to the smooth phase space, when $q\to\infty$, $d\to 1$, ${\mathcal
T}(q)\to 0$, one obtains the statistical weight
\begin{equation}
W_\infty={\rm e}^{\Sigma_\infty},\quad \Sigma_\infty\equiv\sigma_\infty
N\label{24}
\end{equation}
which is determined by the specific complexity $\sigma_\infty$ per one
particle. At small deviation off the minimum complexity $(\Sigma
-\Sigma_\infty\ll\Sigma_\infty)$ and light multifractality
($1-\tau(q)\ll\Sigma_\infty^{-1}$), linearized equation (\ref{22}) gives
\begin{equation}\label{25}
\begin{split}
&W_q\simeq
W_{\tau(q)}\left\{1+\tau(q)\left[1-\frac{2}{3}\left(1-\tau(q)\right)\Sigma_\infty
\right]\left(\Sigma -\Sigma_\infty\right)\right\},\\
&W_{\tau(q)}\equiv\exp\left\{\tau(q)\Sigma_\infty\left[1-\frac{1}{3}\left(1-\tau(q)\right)
\Sigma_\infty\right]\right\},\quad 1\ll q<\infty.
\end{split}
\end{equation}
In the opposite case $\tau(q)\ll 1$, one has with logarithmic accuracy
\begin{equation}
W_q\simeq\left[\alpha\tau(q)\Sigma^2\right]^{\tau(q)}, \ \alpha\sim
1,\quad q-1\ll 1. \label{24a}
\end{equation}


\section{Conclusion}

As shows above consideration, the thermostatistics of complex systems with
phase space, whose reduced fractal dimension $d$ is distributed with spectrum
$f(d)$, is governed by the Tsallis' formalism of the non-extensive statistics.
In this way, the role of non-additivity parameter plays inverted value of the
multifractal function $\tau(q)= qd(q)-f(d(q))$ which monotonically increases,
taking value $\tau=0$ at $q=1$ and $\tau\simeq 1$ at $q\to\infty$ (the latter
limit is related to the smooth phase space). The multifractal function
$\tau(q)$ is reduced to the specific heat to determine, together with the
inverted value ${\bar\tau}(q)\equiv 1/\tau(q)>1$, both statistical
distributions and thermodynamic functions of the system under consideration. At
given function $\tau(q)$, optimization of the normalized multifractal spectrum
$f(d)$ arrives at the dependence of the statistical weight on the system
complexity.

\end{document}